\documentclass[final, 5p, times, twocolumn] {elsarticle}
\usepackage{geometry}
\usepackage{amsmath}
\usepackage{siunitx}
\usepackage{caption}
\usepackage{bm}
\usepackage{textcomp, gensymb}
\usepackage{hyperref}
\hypersetup{colorlinks = true, pdfauthor=author}
\biboptions{sort&compress}
\begin{document}

\captionsetup[figure]{labelfont={default},labelformat={default},labelsep=period,name={Fig.}}
\renewcommand{\figureautorefname}{Fig.}
\begin{frontmatter}
\title{Bi-functional metamaterial based on Helmholtz resonators for sound and heat insulation}
\author[label1]{Brahim Lemkalli \corref{cor1}}
\ead{brahim.lemkalli@edu.umi.ac.ma}
\author[label2]{Zine El Abiddine Fellah}
\author[label3,label4]{S\'{e}bastien Guenneau}
\author[label1]{Khalid Lamzoud}
\author[label5]{Abdellah Mir}
\author[label1]{Younes Achaoui}
\cortext[cor1]{Corresponding author}
\address[label1]{Laboratory of Optics, Information Processing, Mechanics, Energetics and Electronics, Department of Physics, Moulay Ismail University, B.P. 11201, Zitoune, Meknes, Morocco}
\address[label2]{Laboratoire de M\'{e}canique et d'Acoustique, LMA-UMR 7031 Aix-Marseille University-CNRS-Centrale Marseille, F-13453 Marseille CEDEX 13, France}
\address[label3]{The Blackett Laboratory, Physics Department, Imperial College London, SW7~2AZ, London, UK}
\address[label4]{UMI 2004 Abraham de Moivre-CNRS, Imperial College London, SW7~2AZ, London, UK}
\address[label5]{Department of Physics, Moulay Ismail University, B.P. 11201, Zitoune, Meknes, Morocco}
\begin{abstract}

Over the last few decades, both heat and broadband sound reduction have become increasingly significant as a result of concerns about the environment and noise pollution. In order to address this challenge, we provide a finite element analysis study of an acoustic metamaterial panel consisting of a unit cell made of two Helmholtz Resonators with a guide in between. These panels can attenuate and control both sound propagation and heat flux. For noise pollution in a building, we first determine the geometric dimension that corresponds to the operative frequency range. Furthermore, we investigate the sound transmission loss of the proposed panel as a function of the periodicity of an array of unit cells. Additionally, we investigate the thermal flux induced by the panel, especially within a $24$-hour period. The simulation results show that the proposed panel provides a level of sound attenuation within the frequency range of $400$ \si{Hz} to $2.5$ \si{kHz}, as well as interesting heat protection. The structure is compared to panels made up of a homogeneous medium and without Helmholtz resonators.
 
\end{abstract}
\end{frontmatter}

\section{Introduction}

Climate change and noise pollution have emerged as major global issues in recent decades \cite{matos2022innovation,thompson2022noise}. This has generated acoustic and thermal insulation challenges, which usually raise the issue of comfort inside buildings \cite{aly2021acoustic, buratti2018rice}. This also results in a substantial increase in energy consumption, which induces further reliance on construction materials \cite{nurzynski2015thermal}. Different methods based on common building material applications have been examined in order to optimize overall acoustic and thermal performance \cite{dong2023wall}. These materials include organic, inorganic, transparent, and bio-insulation materials, as well as recycled materials \cite{pokorny2022bio, patnaik2015thermal, gao2019improvement, binici2016mechanical, dong2023wall, fornes2022improvement}. Based on the aforementioned materials, it is incredibly difficult to maximize both the acoustic and thermal performances of a building at the same time. This is due to the multiple physical processes associated with these two fields of physics.

Recent advances have proven that artificial man-made materials, called "metamaterials, can successfully control sound wave and heat flux propagation. The optimization of these structures can provide novel functionalities based on distinctive physical properties not found in nature \cite{fan2021review, han2014full, lim2022photonic}. Thanks to their recent advances, metamaterials have recently sparked the interest of academics all over the world. 

Acoustic metamaterials have provided unprecedented noise reduction solutions, especially with regards to mitigating noise propagation. Several variants of metamaterials, such as membranes \cite{mei2012dark, yang2015subwavelength, zhang2020light, jang2022lightweight}, space coiling \cite{li2016acoustic, cai2014ultrathin}, Fabry-Perot structures \cite{wu2000profiled}, quarter-wavelength resonators \cite{shen2021acoustic}, and Helmholtz resonators \cite{li2016sound, jimenez2016ultra, jimenez2017rainbow}, have different characteristics and advantages to enhance sound attenuation. Besides, thermal metamaterials have demonstrated that heat flow may be controlled at will through artificial composite structures \cite{han2014full, li2021transforming}. Heat flux manipulation can be divided into two categories based on the geometric size of thermal metamaterials. The first is that the heat flow could be macroscopically steered throughout thermodynamic conversion, which has been shown to be an effective way to manage heat energy \cite{vemuri2013geometrical, schittny2013experiments}. The second is that nanoscale structures can control the wavelength of the phonons propagation \cite{honarvar2018two, honarvar2016thermal}. Besides the previously described acoustic and thermal metamaterials, the concept of multifunctional metamaterials, which can be used in several physical domains, was also recently proposed \cite{li2020dual, zhang2022dual, fan2023structural}. 

\begin{figure*}
    \centering
    \includegraphics[width=14cm]{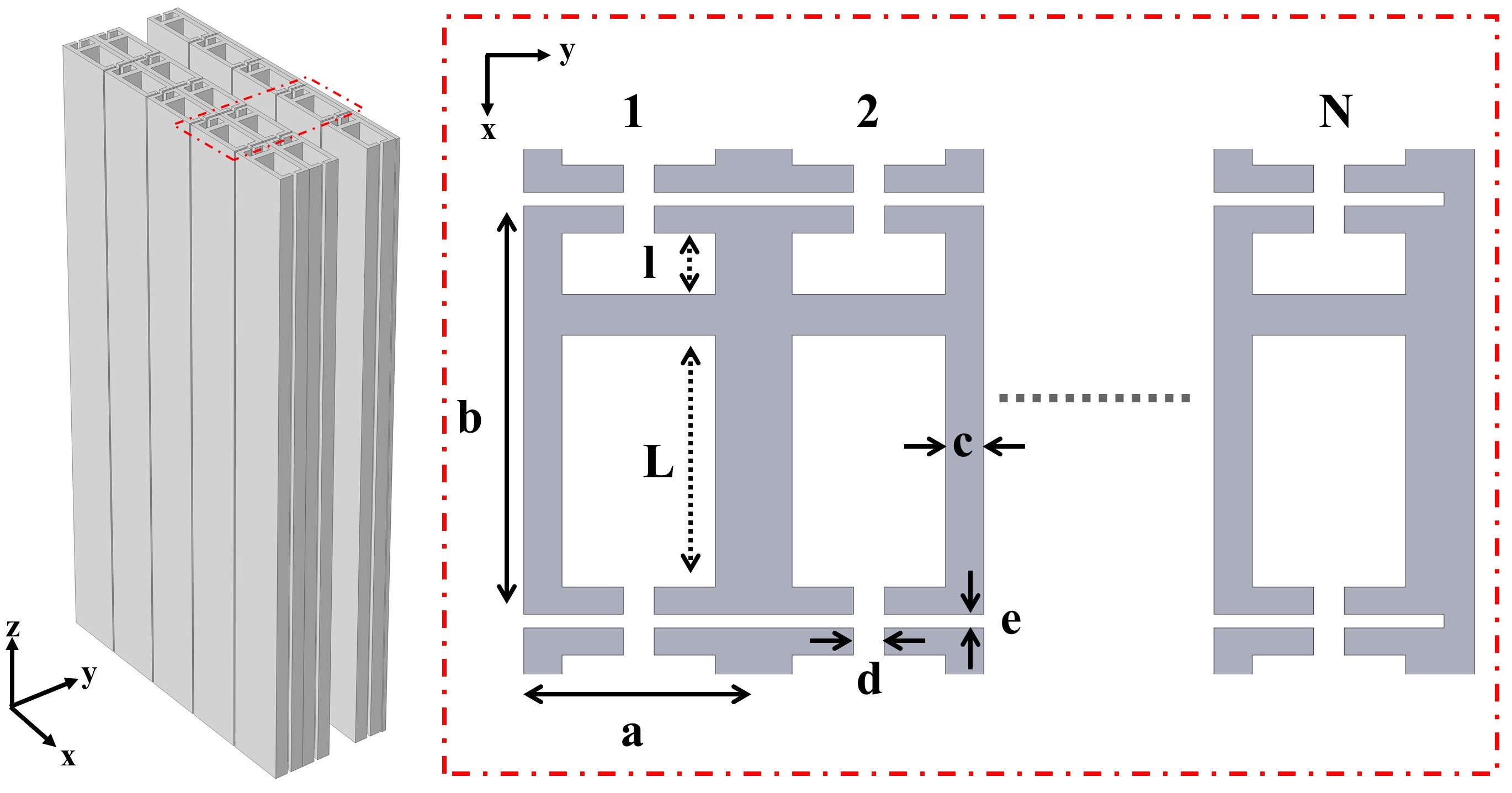}
    \caption{Schematic details of metamaterial based on Helmholtz resonators, including a 3D panel (left) and a cross section in the $xy$-plane (right). The periodicity constants are $a$ along the $y$-axis and $(b+e)$ along the $x$-axis, with $b=2a$ and $e=0.06a$. $N$ is the number of cells along the $y$-axis. The neck size is $d=0.13a$, the length of the small HR is $l=0.33a$, the length of the enormous HR is $L=1.16a$, and $c =0.16a$.}
    \label{Figure 1}
\end{figure*}
In this paper, we propose to use Helmholtz resonators (HRs), which are highly classic acoustic systems that have been intensively explored for decades \cite{rayleigh1916theory}, in order to attenuate sound propagation and confine heat spreading. We suggest panels composed of two symmetrical HRs with periodicity constants ; $a$ along the $y$-direction and $(b+e)$ along the $x$-direction, where $N$ is the cell number corresponding to the $y$-direction. The schematic of the acoustic metamaterial is shown in \autoref{Figure 1}. Each unit cell is made up of two Helmholtz resonators with two square-shaped necks, as illustrated in the exploded view. We analyze the acoustic and thermal behaviors of these panels using the finite element method through two studies: the first aims to determine the sound transmission loss response, and the second involves a temperature analysis that entails evaluating the thermal resistance of these HRs-based panels. We compare the acoustic and thermal properties of the proposed panels to classical plates composed of homogeneous medium and hollowed panels, i.e., HRs without necks.

\section{Numerical Methods}
Numerical models of the acoustic and thermal behaviors of panels are based on 2D calculations. This is due to the symmetry of HRs-based panels in the $z$-direction. In this context, we conduct two studies: the sound transmission loss at low frequencies and the thermal resistance within a $24$-hour period, taking behavior properties numerically within the variation of the number of cells ($N$) in the 2D panel from one to five, and comparing the responses of these panels based on HRs with classical panels as well as hollowed panels (closed cavities). In \autoref{Tab1}, we describe the mechanical and thermal properties of the materials used in computations.
\begin{table}[h]
\caption{\label{Tab1} Mechanical and thermal material properties}
\centering
\begin{tabular}{c c c}\hline
 &Concrete &\\\hline
 Young modulus ($E$) && $25\times10^9$ \si{Pa}\\
 Poisson's ratio ($\nu$)& & $0.2$\\
 Density ($\rho_{s}$) && $2300$ \si{kg/m^3}\\
 Thermal conductivity ($k_s$)&& $1.8$ \si{W/(m K)}\\
 Specific heat ($c_s$)& & $880$ \si{J/(kg K)}\\\hline
 &Air&\\ \hline
 Sound speed ($C$)&& $343$ \si{m/s}\\
 Density ($\rho_{a}$)&& $1.21$ \si{kg/m^3}\\
 Thermal conductivity ($k_{a}$)&& $0.025$ \si{W/(m K)}\\
 Specific heat ($c_{a}$)&& $1005.3$ \si{J/(kg K)}\\\hline
\end{tabular}
\end{table}

A multiphysics model is used to solve the acoustic wave propagation equations in air and solids for the sound transmission analysis. Sound waves in air are governed by the equation \ref{eq001} for the differential pressure $p$. The solid is considered an isotropic linear elastic material, and the acoustic wave propagation is determined by solving the weak formalism of the time-harmonic Navier equation \ref{eq002}. Air-solid interface boundary conditions satisfy the equations corresponding to interface type in the multiphysics coupling; for instance, for the air-solid interface, satisfy equation \ref{eq003}, and for the solid-air interface, satisfy equation \ref{eq004}, where solid tractions equal the pressure \cite{yoon2007topology}.

\begin{equation}\label{eq001}
    \nabla . (\frac{1}{\rho_{a}}\nabla p) + \frac{\omega^2}{\rho_{a} C^2} p=0,\; in\; air
\end{equation}
\begin{equation}\label{eq002}
    \nabla. \sigma=-\omega^2\rho_s\textbf{u}, \;in\; solid
\end{equation}
\begin{equation}\label{eq003}
    n\frac{1}{\rho_a}\nabla p=n\textbf{u}, \;in \;air-solid\; interface
\end{equation}
\begin{equation}\label{eq004}
    n\sigma =np, \;in\; solid-air\; interface
\end{equation}

where $p$ is the acoustic pressure, $\textbf{u}=(u_x, u_y)$ is the displacement vector, $\rho_{a}$ and $\rho_s$ are the densities of the air and solid, respectively. $\sigma=\frac{E}{2(1+\nu)}(\frac{1}{(1-2\nu)} \nabla.\textbf{u}+\nabla \textbf{u})$ is the stress tensor for the isotropic linear elastic material, $C$ is the sound speed in the air, $\omega$ is the angular frequency of the acoustic wave, $E$ is the Young's modulus, $\nu$ is the Poisson's ratio, and $n$ is the outward unit normal to the air or solid.

\begin{figure}
    \centering
    \includegraphics[width=8cm]{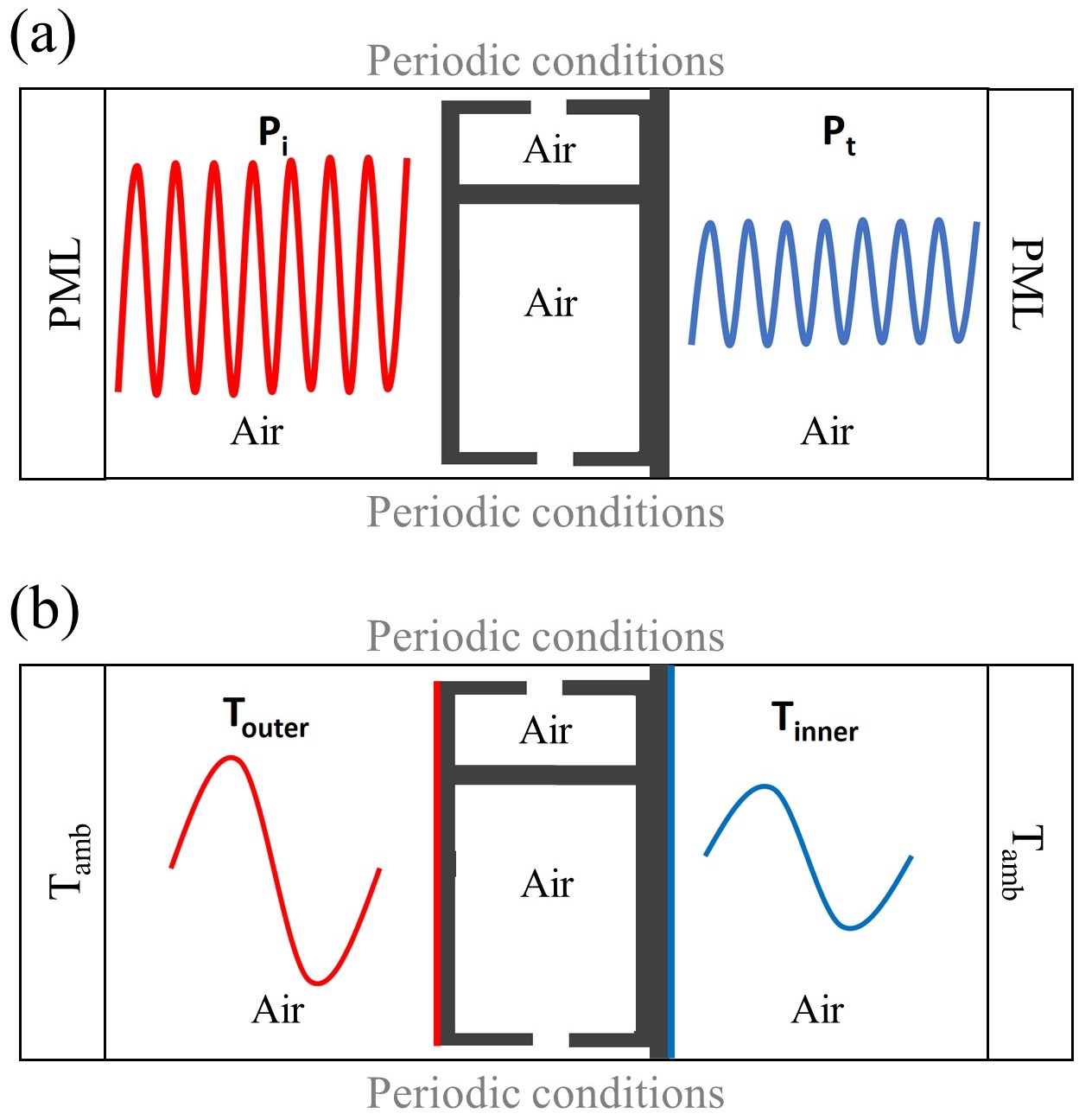}
    \caption{Numerical models. (a) The acoustic model consists of exciting the unit cells with a plane acoustic wave in the frequency range of $20$ \si{Hz} to $10$ \si{kHz}. The transmission is then calculated as the ratio of transmitted pressure to incident pressure. (b) The thermal model involves subjecting the unit cells to a sinusoidal outer temperature $T_{outer}$ and measuring the inner temperature $T_{inner}$ during a $24$ \si{h}.}
    \label{Figure 2}
\end{figure}

In order to simulate the normal incident sound wave, a plane wave radiation boundary condition was utilized, and Perfectly Matched Layers (PMLs) were placed at both extremities of the medium to limit spurious reflections, as depicted in \autoref{Figure 2}(a). Furthermore, the Floquet-Bloch periodic condition is applied in the $x$-direction. We calculated the sound transmission loss (STL) in the frequency range $20$ \si{Hz} to $10$ \si{kHz} for each cell with or without HRs and for the homogeneous plates, using the equation (\ref{eq005}).
\begin{equation}\label{eq005}
    STL=-20\times log_{10}\Big(\frac{|p_t|}{|p_i|}\Big),
\end{equation}
where $p_i$ and $p_t$ are the incident and transmitted pressures, respectively.

For the thermal analysis, the heat transfer in solids and fluids model was utilized to characterize the thermal insulation through the panels. By solving the transient heat equation (\ref{eq006}), the temperature distribution within the unit cells is calculated. The methodology is based on applying a sinusoidal temperature within a $24$-hour period ranging between $0$ \si{\degree C} and $70$ \si{\degree C} and determining the inner temperature of unit cells, as shown in \autoref{Figure 2}(b). To accomplish this, the terminal boundaries of the configuration are exposed to ambient temperature, which is set to $T_{amb}=25$ \si{\degree C}. The outer boundary of cells is vulnerable to the outer temperature $T_{out}(t)$, which describes the temperature variation over a $24$-hour period with the expression $T_{out}=35 sin(\frac{2\pi}{24}t)+35$ \si{\degree C}. Furthermore, the periodicity conditions are applied in the $x$-direction.
\begin{equation}\label{eq006}
    k_i\Delta T=\rho_i c_i \frac{\partial T}{\partial t},
\end{equation}
where $i$ indicates air or solid, $k_i$, $\rho_i$ and $c_i$ are the thermal conductivity, the density, and the specific heat, respectively. 

\begin{figure}[h]
    \centering
    \includegraphics[width=8.5cm]{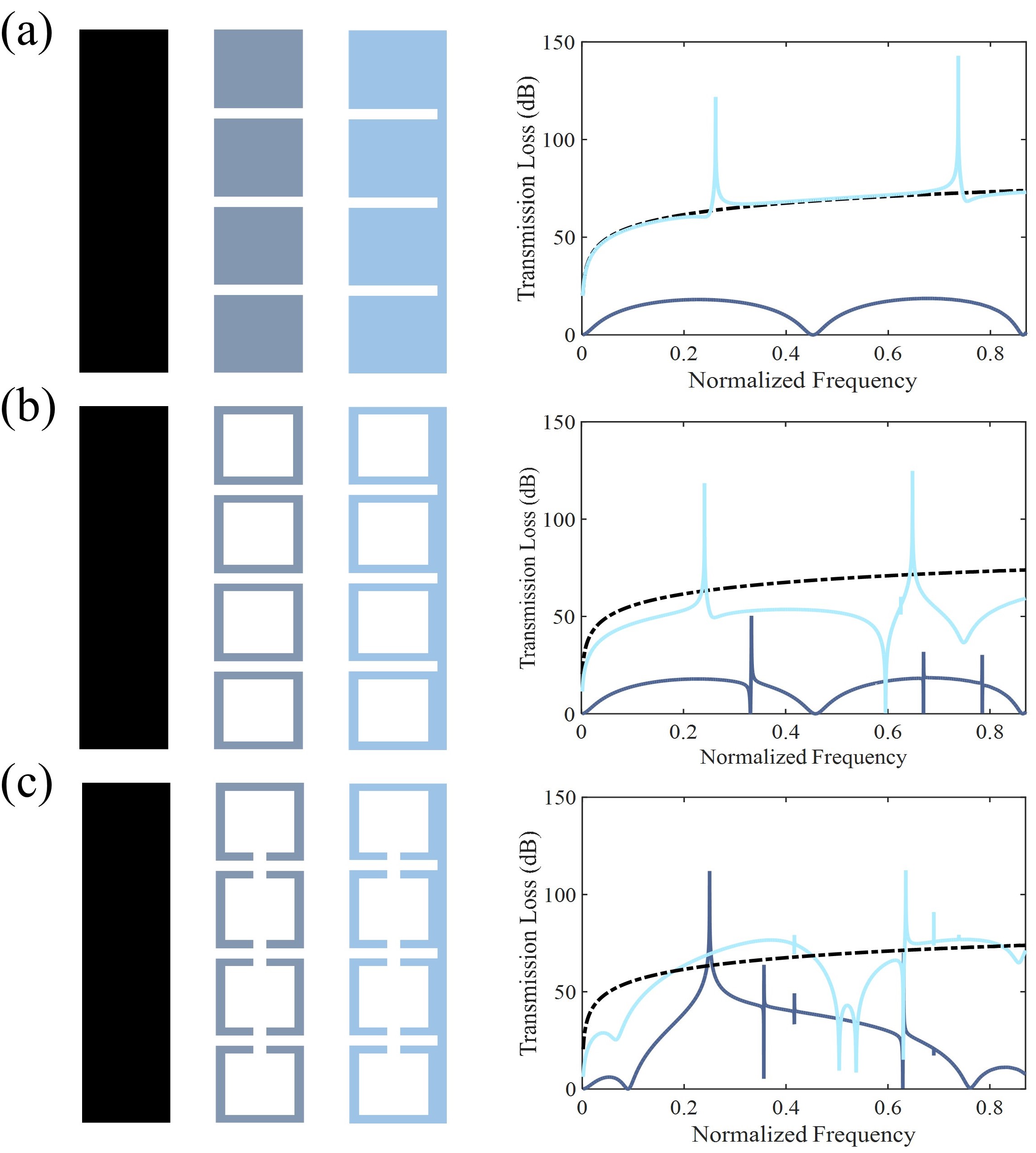}
    \caption{The transmission loss of several configurations is as follows: homogeneous medium on the left, Fabry-Perot in the middle, and quarter-wavelength on the right. (a) Without a cavity. (b) With a cavity. (c) With HRs.}
    \label{Figure 3}
\end{figure}
\begin{figure*}
    \centering
    \includegraphics[width=14cm]{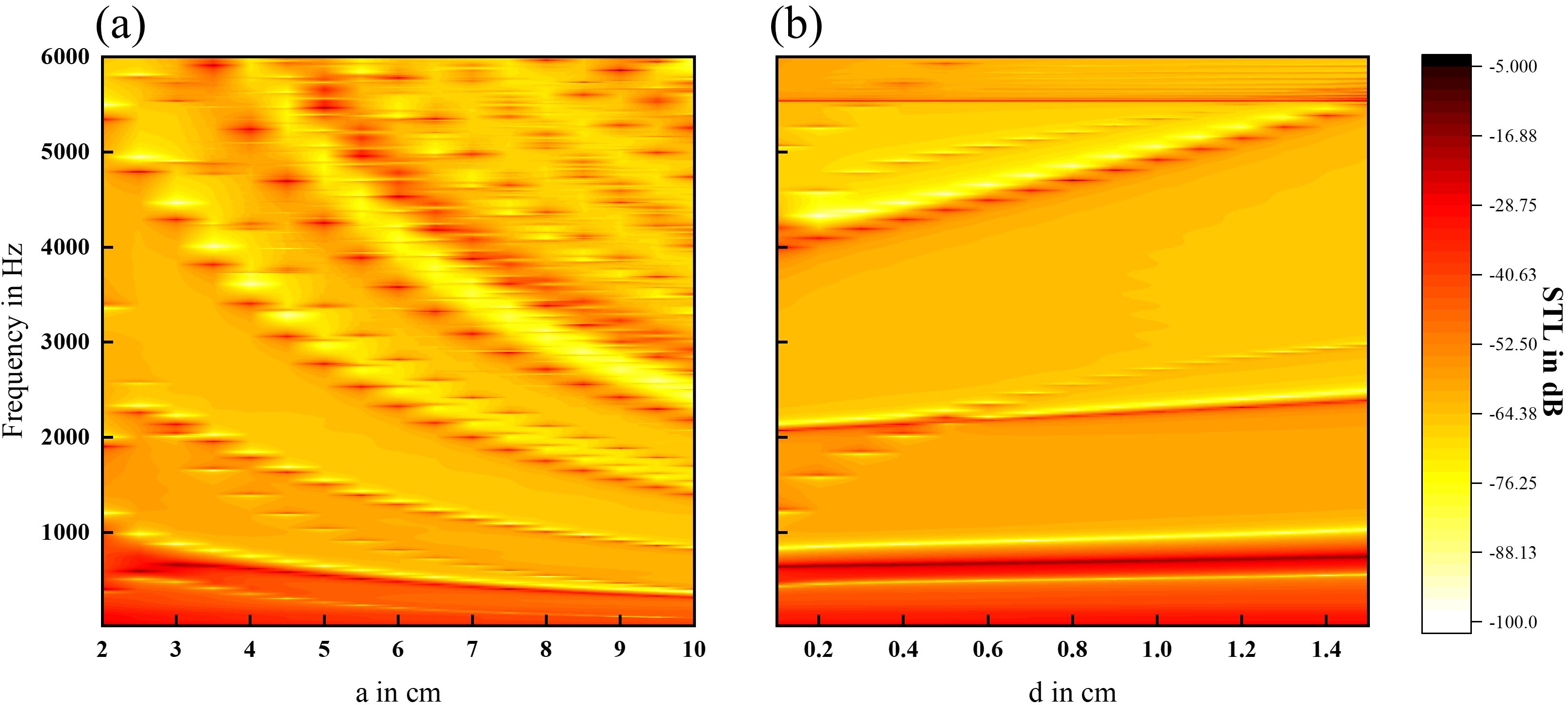}
    \caption{Sound transmission as a function of geometric parameters. (a) The periodicity $a$ size variation. (b) The neck size variation.}
    \label{Figure 4}
\end{figure*}
To gain a better understanding of the resonator mechanism utilized in the panel's design, we compared the transmission loss of various acoustic resonators to the mass law. \autoref{Figure 3} shows the geometrical designs on the left and their corresponding transmission losses on the right. \autoref{Figure 3}(a) depicts the responses of three configurations: homogeneous medium, Fabry-Perot resonator, and quarter-wavelength. It is evident that the Fabry-Perot resonator exhibits resonance frequencies at $n\lambda/2$, and lower transmission loss than the mass law. On the other hand, the quarter-wavelength structure has two resonance peaks with high transmission loss, albeit comparable to the mass law. In \autoref{Figure 3}(b), the main modification to the geometries is the incorporation of cavities into the homogeneous medium. It is evident that the introduction of holes led to the appearance of some resonance peaks. However, the overall impact was not significant when compared to the mass law for the Fabry-Perot and quarter-wavelength resonators. Moreover, the addition of Helmholtz resonators (HRs) to the structures altered the transmission loss curves. The HRs with an open guide exhibited lower transmission loss than the mass law. However, as depicted in \autoref{Figure 3}(c), the combination of HRs and the quarter-wavelength structure led to higher transmission loss than the mass law over a broad range of normalized frequencies. These findings suggest that incorporating HRs into acoustic panels can enhance their sound absorption properties. Additionally, the selection of the resonator geometry was demonstrated to have a significant impact on the effectiveness of sound transmission loss. After identifying the optimal resonator design, we proceeded to determine the geometrical characteristics of the acoustic panel shown in \autoref{Figure 1}.

To promote the use of resonators in the low-frequency range, it is imperative to determine the geometric parameters that allow for the operation of such resonators. This is before we can examine the performance of the insulation in the panels. In the following step, we calculated the sound transmission loss for one unit cell ($N=1$) and varied the periodicity constant $a$ from $2$ \si{cm} to $10$ \si{cm} and neck size $d$ from $0.1$ \si{cm} to $1.5$ \si{cm}. Then the sound transmission loss was calculated in the frequency range from $20$ \si{Hz} to $6$ \si{kHz}. The results of the study are illustrated in \autoref{Figure 4}.

It was found that as $a$ increases, the bandwidth decreases, and thus the diffraction limit descends in frequency (\autoref{Figure 4}(a)). As a consequence, the selected periodicity $a$ has a significant impact on acoustic wave attenuation. The influence of neck size on the two resonators was investigated (\autoref{Figure 4}(b)); the findings reveal that as the neck size increases, the width of the band gap stays invariant and attenuation begins to assume significant values. Therefore, we have selected a periodicity of $3$ \si{cm} and a neck size of $0.4$ \si {cm} for which the diffraction limit begins at $4$ \si{kHz}. This covers a frequency range from $20$ \si{Hz} to $4$ \si{kHz}, i.e. the same frequency range employed in building sound insulation. 

\section{Results and discussion}
\subsection{Acoustic characteristics}
\begin{figure}[h]
    \centering
    \includegraphics[width=8.5cm]{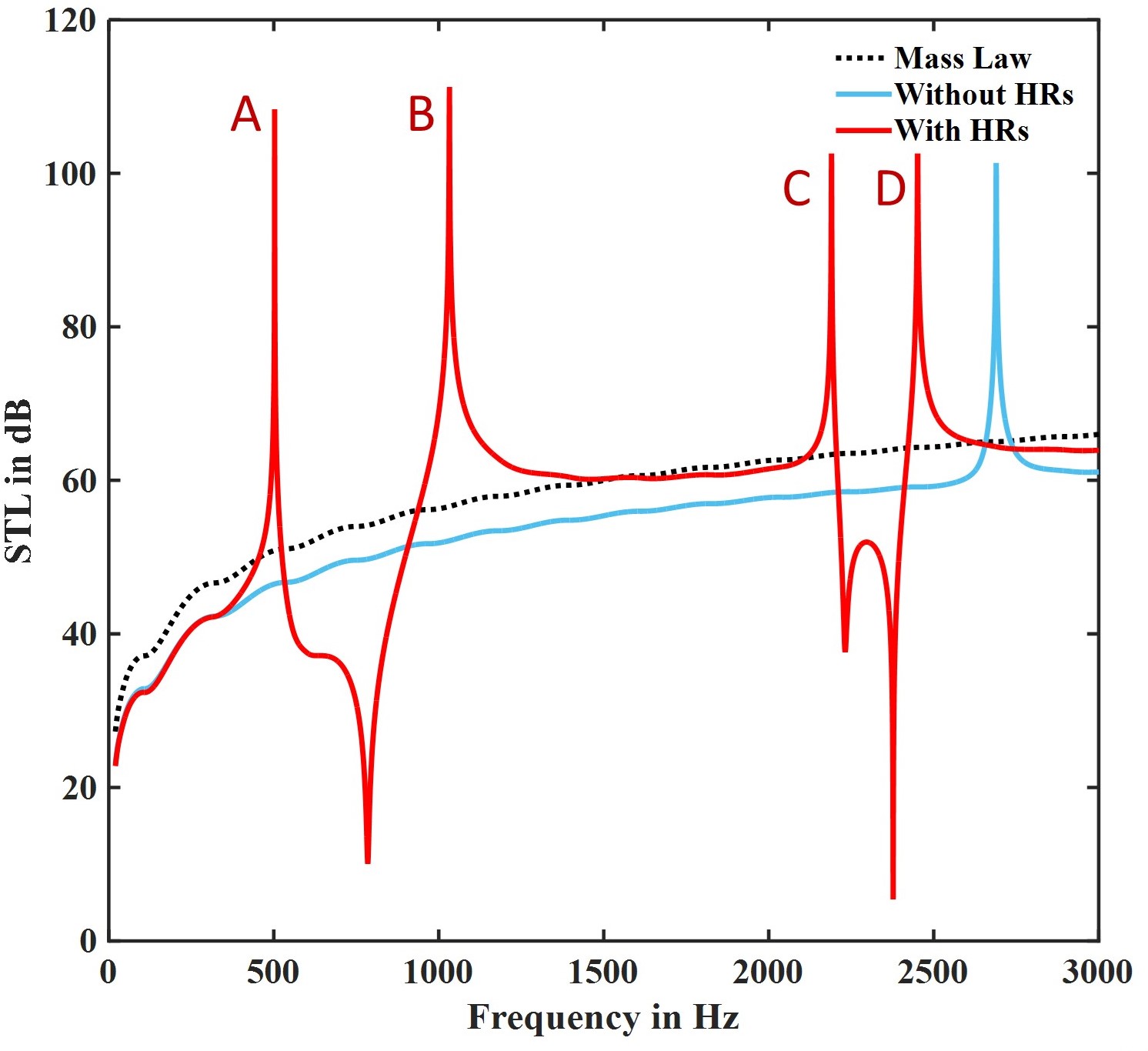}
    \caption{Sound transmission loss for panels with and without HRs in red and blue, and a mass law dotted line in black.}
    \label{Figure 5}
\end{figure}
\autoref{Figure 5} depicts the simulated sound transmission loss (STL) curve for a panel composed of one unit cell with HRs (in red). Moreover, the STL provided by the mass law is presented for comparison (dotted line in black). The operative sound insulation of the panel with HRs defies the mass law in multiple frequency regions: $(418, 533)$ \si{Hz}, $(905, 2211)$ \si{Hz}, and $(2421, 2683)$ \si{Hz}. It assumes that the proposed metamaterial panel can be used to shield low-frequency sound. \autoref{Figure 5} also portrays the panel's STL curve without HRs (in blue). It is worth noting that the STL of a panel without HRs exhibits the same behavior as the mass law in the low-frequency regime $(20, 2500)$ \si{Hz}. As a result, the proposed panel has multiple local resonance mechanisms.

\begin{figure}
    \centering
    \includegraphics[width=8.5cm]{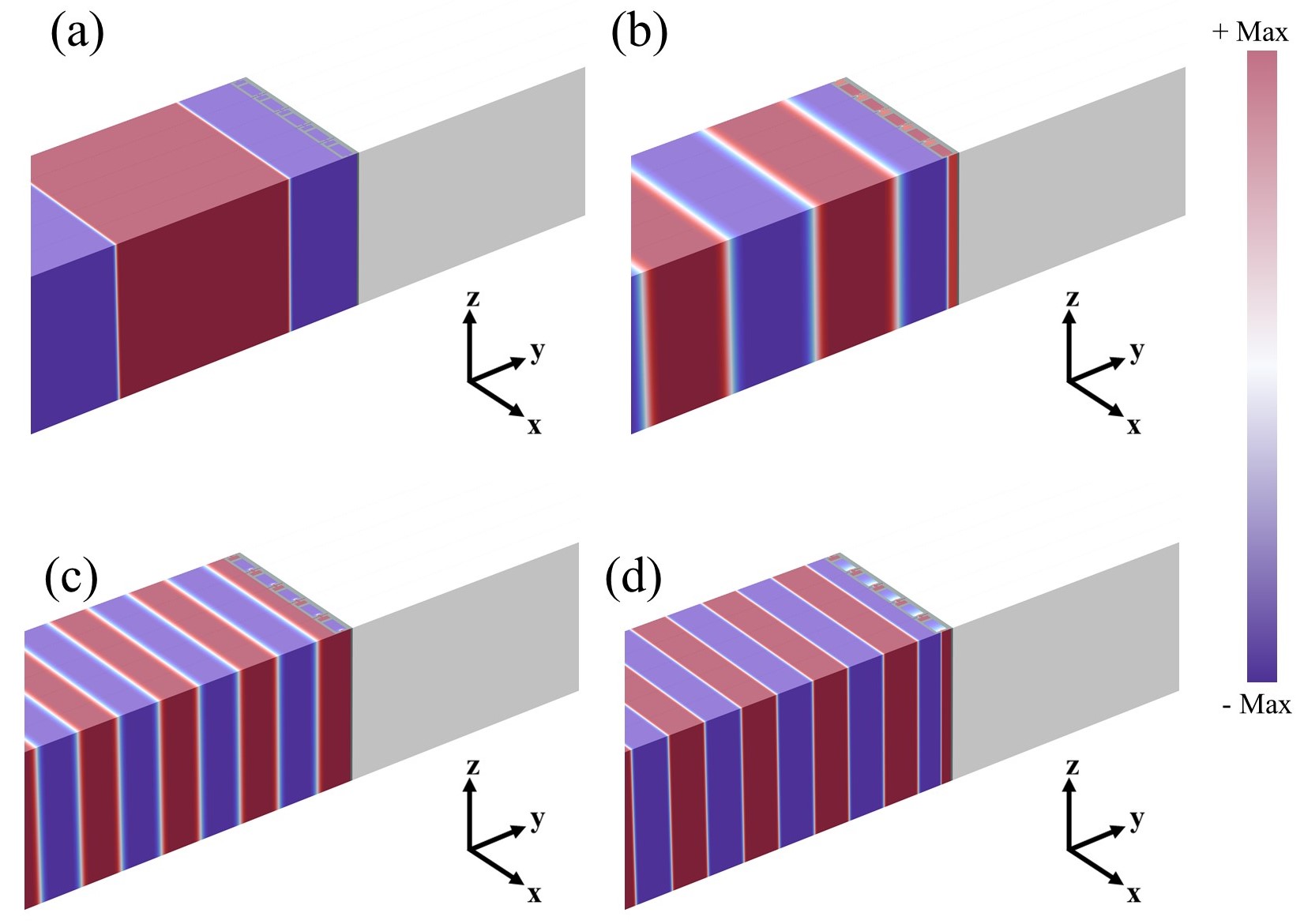}
    \caption{Total pressure screenshots of the panel with HRs. (a) At the frequency of $f_A = 502.6$ \si{Hz}. (b) At the frequency of $f_B = 1033$ \si{Hz}. (c) At the frequency of $f_C = 2190$ \si{Hz}. (d) At the frequency of $f_D = 2451$ \si{Hz}.}
    \label{Figure 6}
\end{figure}
\begin{figure}
    \centering
    \includegraphics[width=8.5cm]{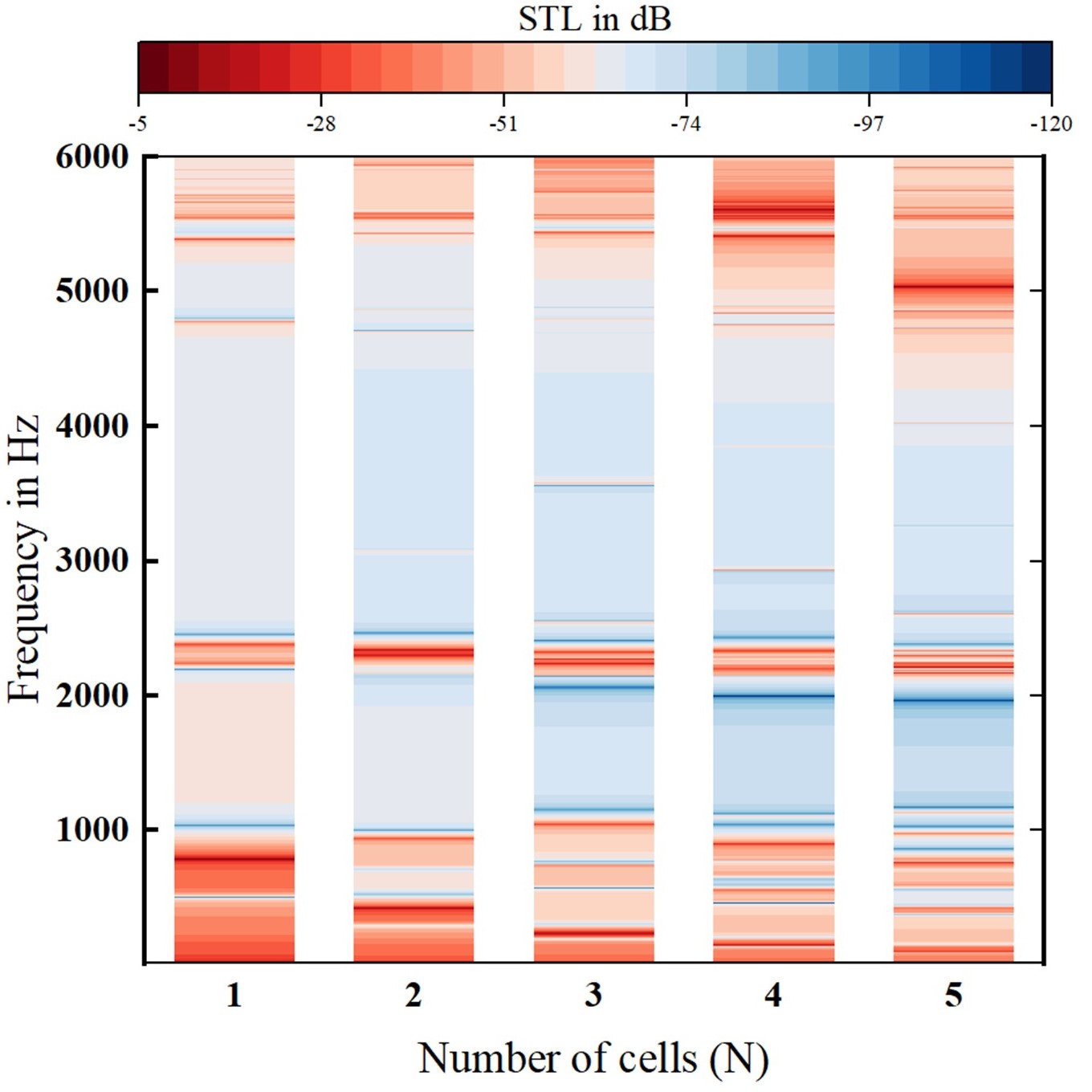}
    \caption{Sound transmission loss as a function of the number of cells $N$ in the panel with HRs.}
    \label{Figure 7}
\end{figure}

\autoref{Figure 6} depicts the total pressure of a panel composed of one unit cell at frequencies $f_A = 502.6$ \si{Hz}, $f_B = 1033$ \si{Hz} $f_C = 2190$ \si{Hz} and $f_D = 2451$ \si{Hz} to gain further insight into the local resonance behaviors. However, within the broad band generated by the existence of HRs, there is a resonance peak corresponding to a single frequency. Since noise is formed by a chirp with a wide frequency range, this peak has no effect on noise insulation. 
\autoref{Figure 7} portrays the STL of the panel made up of $N$ unit cells containing the Helmholtz resonators. As previously discussed, the STL reduces the amplitude of sound transmission loss over a wide frequency range for $N=1$. Moreover, panels with two $N=2$ unit cells have a higher STL and a broader bandwidth than the previous one, whereas panels with three $N=3$ unit cells, four $N=4$ unit cells, and five $N=5$ unit cells all have improved acoustic performance as the number of cells grows. As a consequence, as the number of cells increases, sound transmission reduces.

To summarize, the panel with HRs outperforms its counterpart made of homogeneous medium and lacking HRs in terms of sound insulation. In other words, the numerical results of the acoustic analysis show that cells with Helmholtz resonators have a significantly higher level of sound isolation than cells without Helmholtz resonators. This means that, in the frequency range of $20$ to $3000$ \si{Hz}, HRs confine the incoming sound wave more effectively than ordinary panels.

\subsection{Thermal characteristics}
\begin{figure}
    \centering
    \includegraphics[width=8.5cm]{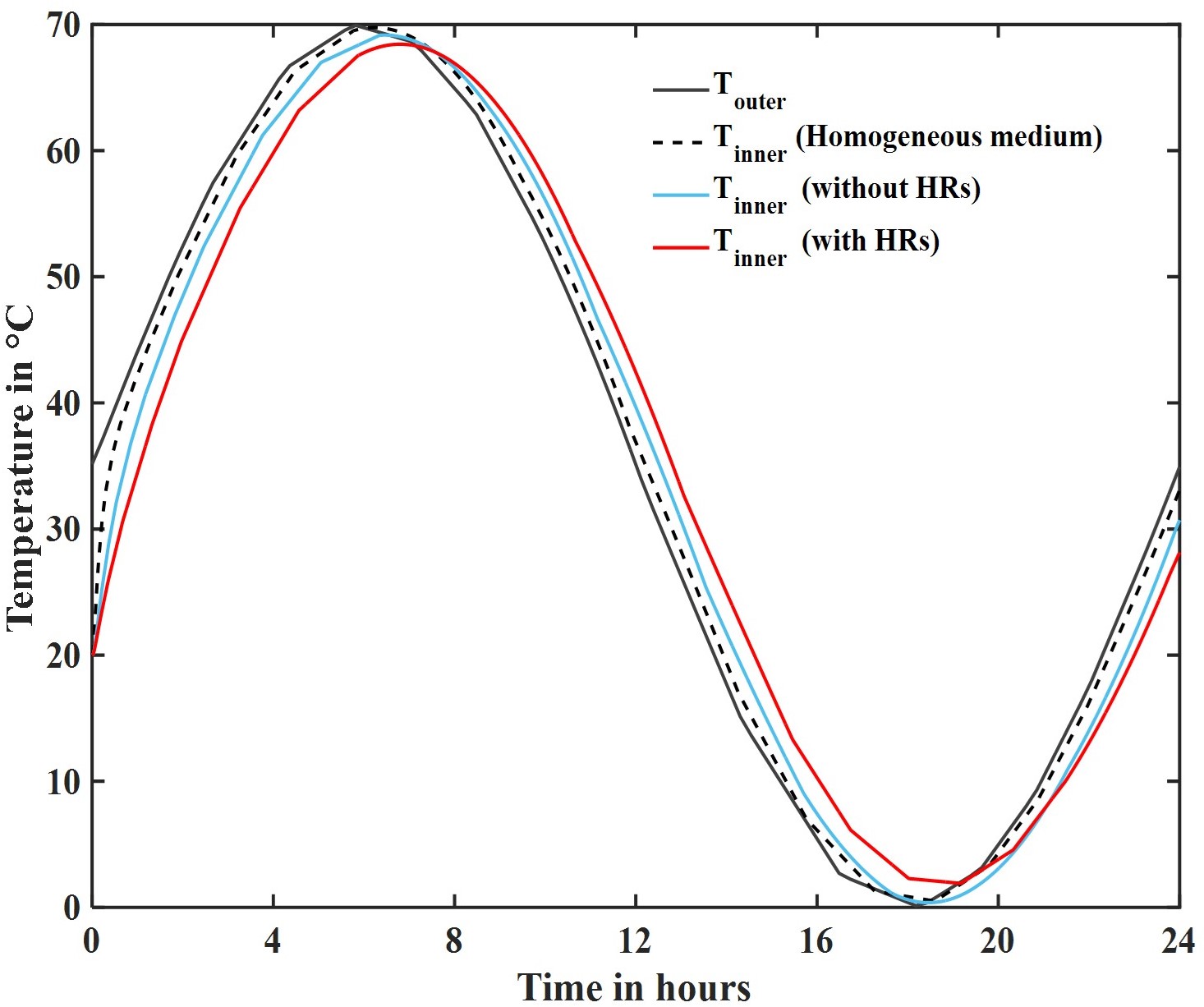}
    \caption{The inner temperature fluctuations as a function of exposure time for panels with and without HRs in red and blue, respectively, and a homogeneous medium dotted line in black.}
    \label{Figure 8}
\end{figure}
\begin{figure}[h]
    \centering
    \includegraphics[width=8.5cm]{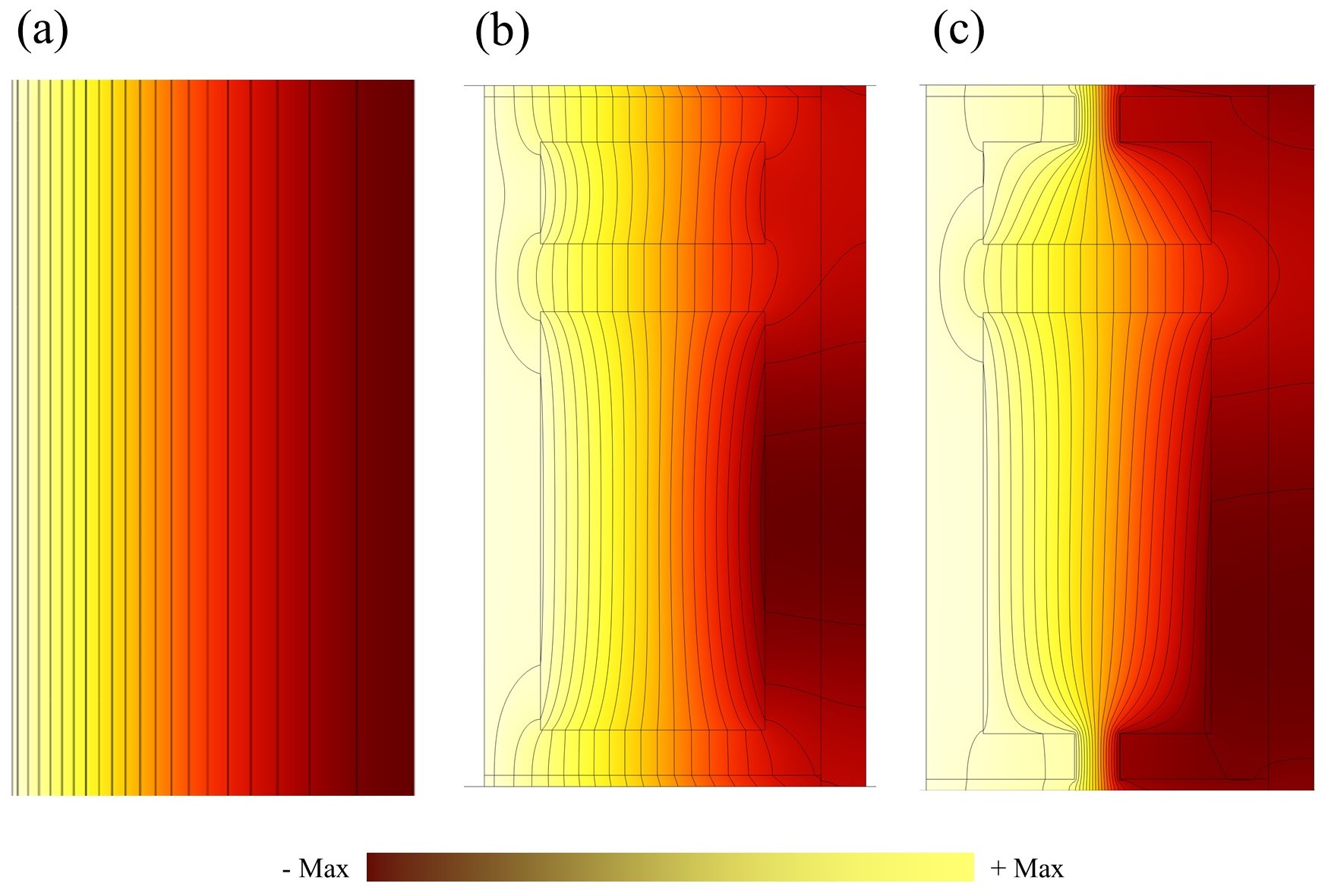}
    \caption{The screenshots of temperature counters after 4 hours of exposure: homogeneous medium (A), a panel without HRs (B), and a panel with HRs (C).}
    \label{Figure 9}
\end{figure}

\autoref{Figure 8} represents the temperature evolution as a function of time for panels composed of one unit cell. For $24$ hours, the panels are exposed to a sinusoidal outer temperature (colored in black) varying between $0$ and $70$ \si{\degree C} on their outer side. The evolution of the inner temperature of the panel with HRs (colored in red) is not the same as the inner temperatures of the other two panels composed of the homogeneous medium (dotted line in black) and without HRs (colored in blue). Thus, the inner temperature changes between the three panels and is therefore not identical. In this case, it is possible to infer that the panels have different thermal resistances. As a result, the panel containing HRs has stronger resilience than the other cells. Indeed, cells containing HRs take longer time to reach the maximum and minimum temperatures than cells without HRs, which is related to the slots presented in the HRs. Screenshots taken after four hours of exposure to the temperature distribution are used to gain a better understanding of the thermal resistance of the three panels, as shown in \autoref{Figure 9}. It is clear that the distribution of the counter temperature is not the same for all three cases; the presence of the HRs makes the panel more resilient and increases its heat transfer mechanism.

\begin{figure}
    \centering
    \includegraphics[width=8.5cm]{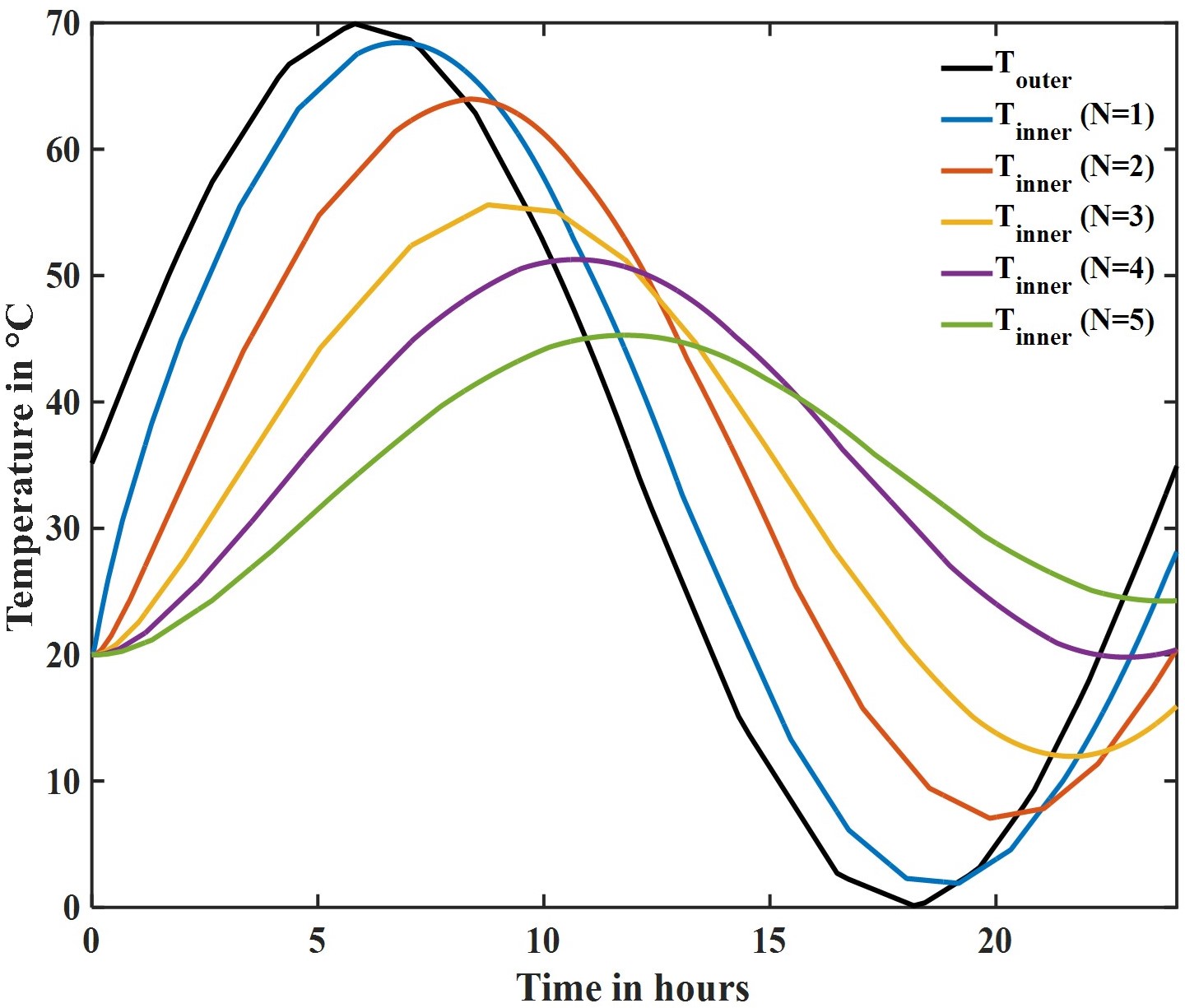}
    \caption{The inner temperature of N unit cells with HRs as a function of time.}
    \label{Figure 10}
\end{figure}

The inner temperature evolution of the panel of $N=2$ becomes more resilient than that of the panel composed of a single unit cell with HRs. Additionally, the variation of the inner temperature as a function of time changes with the number of cells $N=3$, $N=4$, and $N=5$, as illustrated in \autoref{Figure 10}. The thermal resistance of the panels increases with the number of cells in the $y$-direction.

It is noteworthy that the presence of HRs in the panels functioned as barriers to reduce heat propagation. This demonstrates, in particular, that the improved panels provide higher thermal insulation than conventional panels consisting of homogeneous medium or cells without HRs. The presence of Helmholtz resonators in unit cells influences their thermal properties, resulting in extrathermal change processes. As a result, thermal transmission occurs through three complementary mechanisms in the unit cell: conduction through the solid, radiation, and convection by air within and outside the cavities.
\section{Conclusion}

To conclude, we demonstrated a panel based on Helmholtz resonators in order to reduce both sound and heat transmission. A series of simulations were performed using the finite element method to characterize the physical characteristics of the proposed panels in the context of determining their physical characteristics. According to the sound transmission loss analysis performed on the proposed metamaterial, the metamaterial's resonance elements (HRs) perfectly reflect the incoming energy by trapping it by its anti-resonance mechanism in the frequency range corresponding to the noise inside buildings, which ranges in three regions: $(418, 533)$ \si{Hz}, $(905, 2211)$ \si{Hz}, and $(2421, 2683)$ \si{Hz}. The results of the heat transfer analysis of panels with HRs indicate that, in comparison to homogeneous panels and panels without HRs, heat propagation decreases with exposure time. In other words, the presence of HRs mitigates the spread of heat fluxes. Conversely, when Helmholtz resonators are involved, it is as if a thermal barrier is created that prevents thermal fluxes from spreading outward. Thus, we were able to develop a bi-functional metamaterial that may be used in the building sector to alleviate noise and heat concerns.

\section*{Acknowledgements}
We would like to thank the Moroccan Ministry of Higher Education, Scientific Research and Innovation and the OCP Foundation who funded this work through the APRD research program.

\bibliography{Mybib}

\end{document}